\documentclass{PoS}

\usepackage{graphicx}
\usepackage{amssymb}
\usepackage{amsmath,esint}
\usepackage{subfig}

\title{Light diffusion in birefringent polycrystals and the IceCube ice anisotropy}

\ShortTitle{Light diffusion in birefringent polycrystals and the IceCube ice anisotropy}

\author{
The IceCube Collaboration\footnote{For collaboration list, see PoS(ICRC2019) 1177.}\\
{\itshape \href{http://icecube.wisc.edu/collaboration/authors/icrc19_icecube}{http://icecube.wisc.edu/collaboration/authors/icrc19\_icecube}}\\
E-mail: \email{dima@icecube.wisc.edu, martin.rongen@icecube.wisc.edu}
}

\abstract{

The IceCube Neutrino Observatory instruments about 1\,km$^3$ of deep, glacial ice at the geographic South Pole with 5160 photomultipliers to detect Cherenkov light from charged relativistic particles. The experiment pursues a wide range of scientific questions ranging from particle physics such as neutrino oscillations to high-energy neutrino astronomy. Most of these efforts rely heavily on an ever more precise understanding of the optical properties of the instrumented ice. An unexpected light propagation effect, observed by the experiment, is an anisotropic attenuation, which is aligned with the local flow of the ice. The exact cause is still under investigation. In this contribution, the micro-structure of ice as a birefringent polycrystal is explored as the cause for this anisotropy.

\vspace{4mm}
{\bfseries Corresponding authors:}
\speaker{Dmitry Chirkin}$^{1}$, Martin Rongen$^{2}$\\
{$^{1}$ \itshape Dept. of Physics and Wisconsin IceCube Particle Astrophysics Center, University of Wisconsin, Madison, WI 53706, USA}\\
{$^{2}$ \itshape III. Physikalisches Institut, RWTH Aachen University, D-52056 Aachen, Germany}

}

\FullConference{36th International Cosmic Ray Conference -ICRC2019-\\
		July 24th - August 1st, 2019\\
		Madison, WI, U.S.A.}

\begin{document}

\section{Introduction}

IceCube is a cubic-kilometer neutrino  detector installed in the ice at the geographic South Pole     \cite{detector:paper} instrumenting depths between 1450\,m and 2450\,m. Neutrino reconstruction relies on the optical detection of Cherenkov radiation emitted by secondary particles produced in neutrino interactions in the surrounding ice or the nearby bedrock. The optical  properties of ice surrounding the  detector are described with a table of absorption and effective scattering coefficients approximating average ice properties in 10\,m-thick ice layers. These  properties were determined with a dedicated calibration measurement as described in \cite{Aartsen2013}.  Nearly every optical sensor (digital optical module, or DOM) of the detector was operated in ``flasher'' mode (one at a time) to emit light from  on-board LEDs  in an approximately  azimuthally-symmetric pattern. The emitted light was then observed by the DOMs on the surrounding strings.

\begin{figure}[h]
    \centering
    \includegraphics[width=0.65\textwidth]{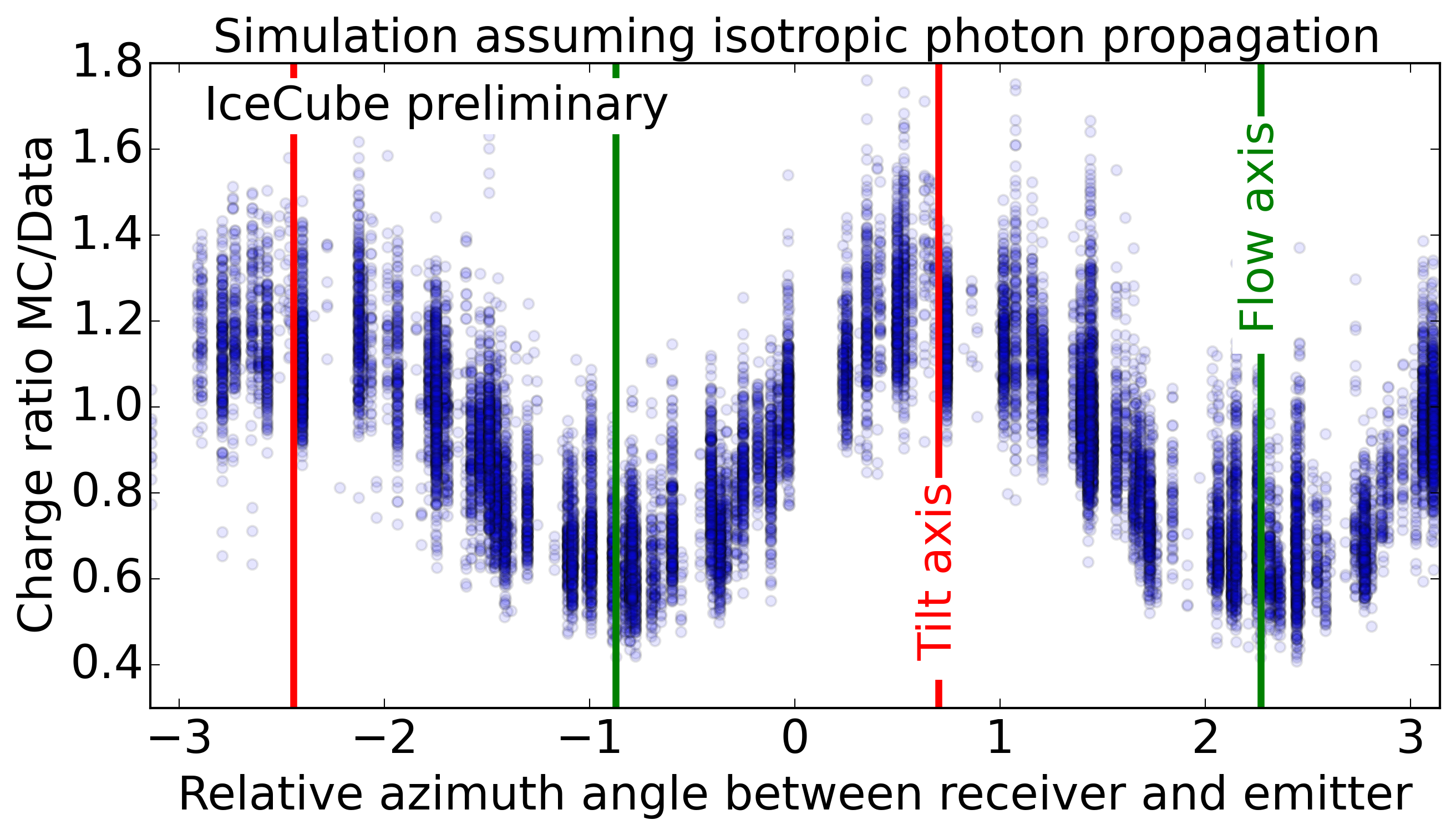}
    \caption{Optical ice anisotropy seen as azimuth dependent charge excess in flasher data.}
    \label{fig:ratio}
\end{figure}

As previously reported  \cite{ICRC_anisotropy}, IceCube observes a strong anisotropy in the light propagation at macroscopic scales. Measured at $\sim$125\,m from an isotropic emitter (averaging over many flashers), about twice as much light reaches DOMs on the flow axis then on the orthogonal tilt axis (see Figure \ref{fig:ratio}). At the same time the arrival time distributions are nearly unchanged.

Several parametrizations modifying the scattering function, absorption and/or scattering coefficients as a function of propagation angle have been explored in the past with some success. However, none of them were able to fit charge and timing distributions simultaneously. Departing from the paradigm that the optical properties are driven by particulate impurities \cite{Aartsen2013}, this report studies an effective photon deflection and a direction-dependent diffusion caused by light birefringence in polycrystals of Antarctic ice. This might be a natural and well motivated mechanism responsible for the large scale optical anisotropy as observed in the IceCube detector.

\section{Birefringence}

Birefringent crystals, such as calcite, are fascinating objects and light diffusion in birefringent, polycrystalline materials has been discussed as early as 1955  \cite{Raman1955}. While the literature agrees that the combined effect of ray splitting on many crystal interfaces will lead to a continuous beam diffusion, the resulting diffusion patterns remain largely unexplored. 

In a homogeneous, transparent and non-magnetic medium the relation between the electric field and the displacement field as well as the magnetic fields is given as \cite{Landau}:
\begin{eqnarray}
\vec{B} = \vec{H}, \hspace{3mm} 
\vec{D}=\epsilon \vec{E}
\end{eqnarray}

As the dielectric tensor $\epsilon$ is symmetric, one can always find a coordinate system where it is diagonal $\epsilon = diag(n_x^2, n_y^2, n_z^2)$, with $n_i$ being the refractive index along the given axis. Uniaxial crystals, such as ice, have two distinct refractive indices: $n_x = n_y \equiv n_o \neq n_z \equiv n_e$. The axis of the different refractive index $n_e$ defines the optical axis (and coincides with c-axis, see Sec.\ \ref{icebire}).

\begin{figure}[h]
    \centering
    \subfloat[Ordinary ray]{\includegraphics[width=0.33\textwidth]{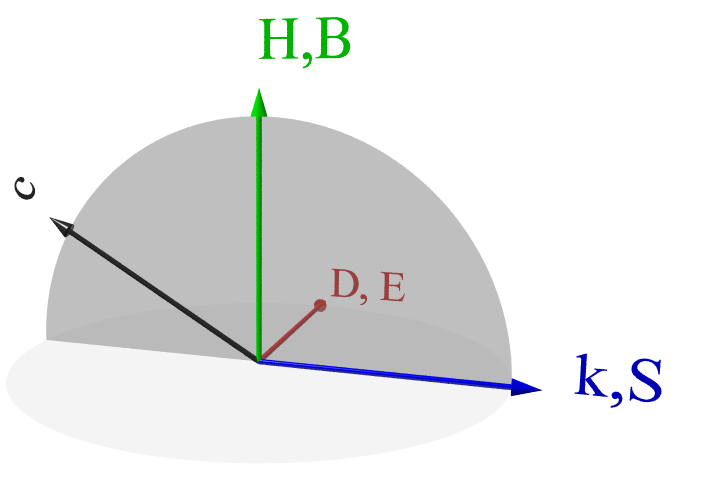}}\hspace{1cm}
    \subfloat[Extraordinary ray]{\includegraphics[width=0.27\textwidth]{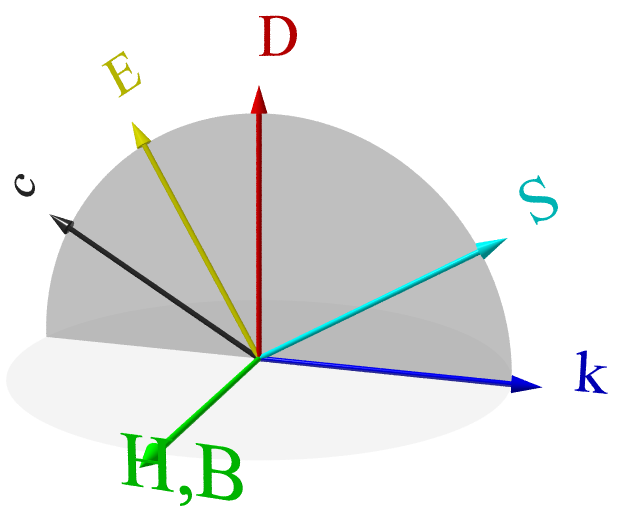}}
    \caption{Orientation of all electromagnetic vectors for the ordinary and extraordinary ray with respect to the crystal axis (c-axis). Adapted from \cite{keller}.}
\end{figure}

Light propagating in a uniaxial crystal is split into an ordinary wave and an extraordinary wave of orthogonal polarizations. The electric field vector $\vec{E}$ and the displacement vector $\vec{D}$ for the ordinary wave are always collinear to each other and perpendicular to both the optical axis of the crystal and the parallel propagation vectors $\vec{k}$ and $\vec{S}$. However, the electric field $\vec{E}$ for the extraordinary wave is not, in general, perpendicular to the propagation vector $\vec{k}$. It lies in the plane formed by the propagation vector and the displacement vector. The electric field vectors of these waves are mutually orthogonal \cite{ZHANG1996549}. The energy flow is given by the Poynting vector $\vec{S}= \frac{c}{4\pi}\vec{E} \times \vec{H}$. For the extraordinary wave, the Poynting vector $S$ is not parallel to $k$.

While the ordinary ray always propagates with the ordinary refractive index $n_o$, the refractive index of the extraordinary ray depends on the opening angle $\theta$ between the optical axis and the wave vector $\vec{k}$ as given in equation \ref{eq:ne}.

\subsection{Ice birefringence}
\label{icebire}

Ice is the comprehensive term for all twelve known solid phases of water. For temperatures and pressures naturally occurring on Earth only the hexagonal crystal form, called Ih, occurs. Each monocrystal consists of a neat stack of layers and each layer can be thought of as a tessellation of hexagonal rings, with oxygen atoms at each node  \cite{Faria2014-part2}. The plane defined by these layers is called the basal plane, with the normal vector referred to as the principal axis or c-axis, as it is the axis of highest rotational symmetry. This c-axis is also the optical axis of the crystal \cite{Faria2014-part2}.

The birefringence strength can be expressed as:

\begin{equation}
    \beta = \left(\frac{n_e}{n_o}\right)^2 -1 .
\end{equation}

For ice  $\beta$ is $\approx 2\cdot 10^{-3}$ across the entire visible wavelength spectrum \cite{Petrenko}.

\subsection{Glacial ice as a deforming polycrystal}

The ice found in glaciers such as on the Antarctic plateau has been created through compactification of snow layers into firn and at a later stage ice. Thus it consists of a large number of interlocked mono-crystals, also called grains. They are typically sub-millimeter to centimeter in size. The surface where two grains meet are called grain boundaries  \cite{Faria2014-part2}.

Ice flows under its own weight, either as a bulk movement down a slope or through plastic deformation. Plastic deformation is mediated through deformations of individual grains as well as interactions between the grains. The viscosity of an individual crystal strongly depends on the direction of the applied strain. A hexagonal crystal will most readily deform as shear is applied orthogonal to the c-axis leading to slip of the individual basal planes. Therefore individual grains elongate, with the direction of elongation being perpendicular to the c-axis  \cite{Weikusat20150347}.

The true three-dimensional c-axis orientation of individual grains can be measured using polarized light microscopy  \cite{Faria2014-part2}. The generally observed evolution of the c-axis distribution versus depths in glacial ice behaves as follows:
In the top third of the glacier c-axis are distributed randomly. This slowly develops into a so called girdle fabric where most all c-axes are found in a plane orthogonal to the flow. At 80\% to 90\% of the total depth a second rapid transition to a single vertical cluster can be observed  \cite{Weikusat20150347}.

\subsection{Parameters for the South Pole ice}

\subsubsection{c-axis distributions}

\begin{figure}[h]
    \centering
    \includegraphics[width=\textwidth]{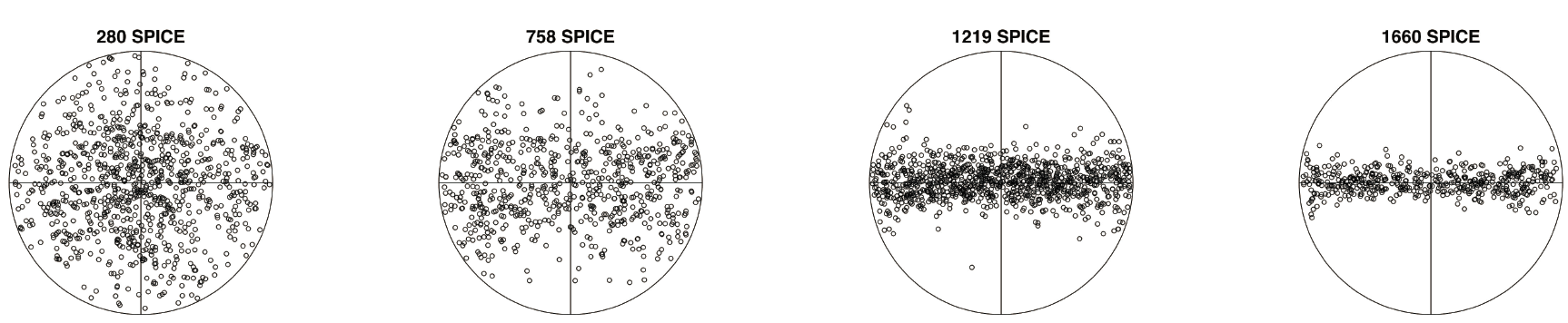}
    \caption{Depth development of c-axis distributions (Lambert azimuthal equal-area projections) measured in SPICEcore; depths (in meters) are indicated atop each diagram. They show a clean girdle fabric at IceCube depths \cite{SpiceCorePoster}.}
    \label{fig:SPICE}
\end{figure}

The samples from the SPICEcore ice core \cite{Casey2014} in close vicinity to IceCube reach 1751\,m depth which corresponds to ice layer depths of $\sim$1820\,m in the IceCube coordinate system. C-axis distributions have been measured at all depths and show an exceptionally clean girdle fabric at depths near the bottom of SPICEcore (see Figure \ref{fig:SPICE}), which overlap with the top third of IceCube-instrumented depths  \cite{SpiceCorePoster}.

\subsubsection{Grain size and elongation}

The grain size distribution, that is the size distribution of ice mono-crystals, defines the distance between interface crossings. Grains are on average elongated, which introduces an additional azimuth/zenith dependence on the effective grain size. Grain size and elongation distributions have not yet been published by the SPICEcore collaoration, but are from other cores expected to be on the mm-scale with elongations of at most a factor of two \cite{Weikusat20150347}. 

\section{Analytic calculation}

Assuming an arbitrary ray incident on a plane interface, we first calculate the four possible wave vectors, the ordinary and extraordinary refracted rays and the ordinary and extraordinary reflected rays. Given the wave vectors, the four associated Poynting vectors are calculated from the boundary conditions, yielding the energy flow and as such probable photon directions.

\subsection{Wave vectors}

\begin{figure}[h]
\centering\includegraphics[width=0.5\linewidth]{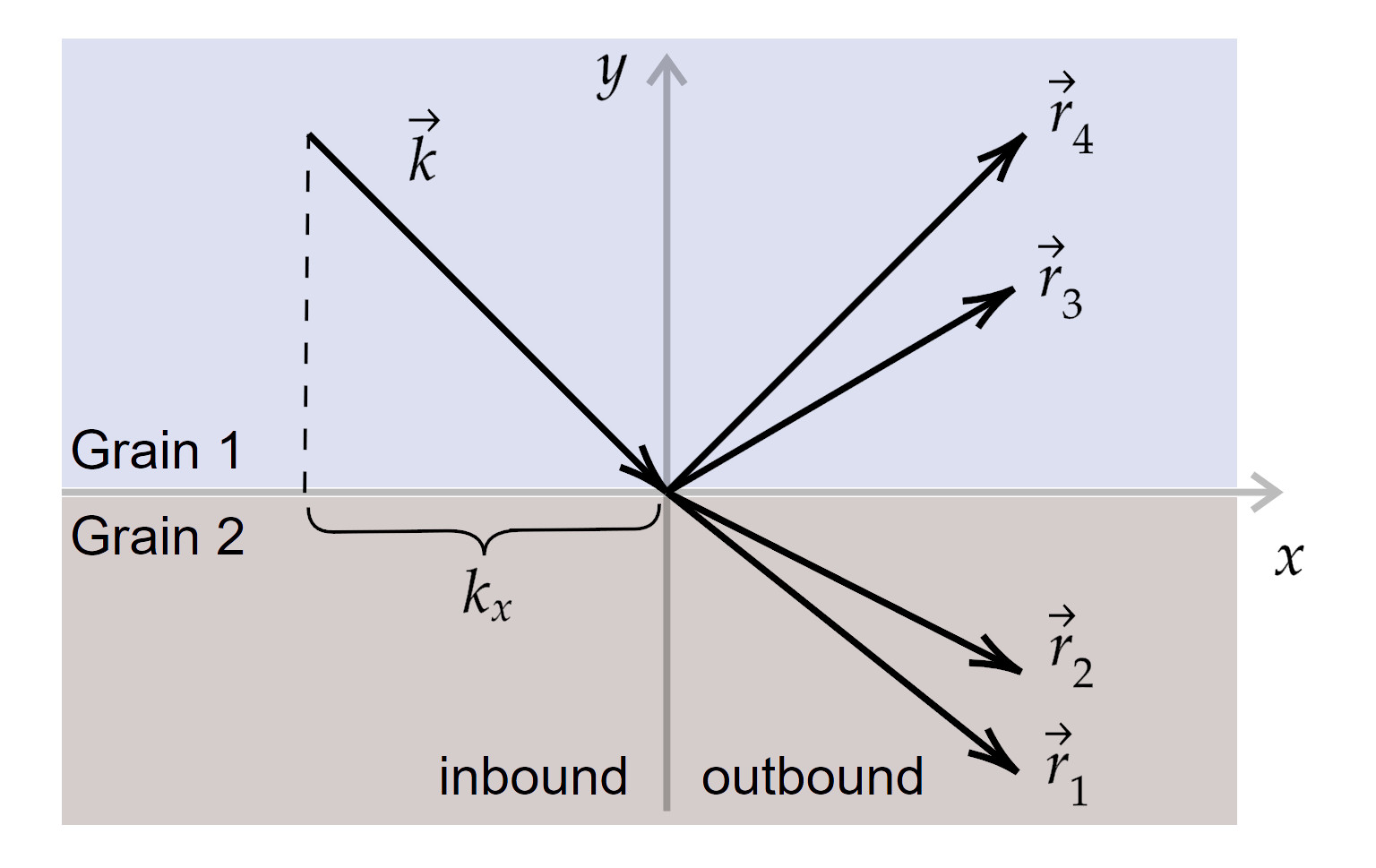}
\caption{Sketch of wave vectors for the incident, reflected and refracted rays. The surface component is always conserved.}
\label{fig:rays}
\end{figure}

Figure \ref{fig:rays} shows the situation at hand. An incoming wave vector $\vec{k}$ intersects the interface and is split into four outgoing wave vectors $\vec{r}$. The coordinate system can always be chosen such that the surface normal $\vec{n}$ is along the y-axis and that the surface components of $\vec{k}$ and as such $\vec{r}$ are along the x-axis. Here we implicitly assume, as an approximation, that the boundary surface is a perfect plane infinite in its extension, and, without a loss of generality, that the incoming and outgoing waves are all plane waves.

Because of translational symmetry of the interface surface, the surface components of all wave vectors are identical  \cite{Landau}: $k_x=r_x$. As the wave number is given by $k=\frac{2\pi}{\lambda}$, we can define a vector $\vec{n}$ such that $\vec{k}=\omega \vec{n}/c$, whose magnitude $n$ is the direction dependent refractive index $n=\sqrt{\epsilon (\theta)}$. As such the magnitude of the wave vector is proportional to the refractive index and we shall simplify $|\vec{k}|=n$ in the following. 

\subsubsection{Outgoing ordinary rays}

Given the magnitude $n_o$ and surface component $k_x$ of the wave vector the y-component is simply calculated as:
\begin{equation}
r_y = \pm \sqrt{n_o^2-k_x^2}
\end{equation}

Given two media with different refractive indices, one obtains Snell's law for refraction and the usual law for reflection ($k_y=r_y$). The outgoing ordinary ray of an inbound ordinary ray is not deflected, as it does not see a change in refractive index.

\subsubsection{Outgoing extraordinary ray}

Determining $r_y$ for the extraordinary rays follows the same logic, only with a refractive index which depends on the opening angles $\theta$ between the refracted wave vector $\vec{r}=(r_x,r_y)$ and the optical axis $\vec{a} = (a_x, a_y, a_z)$:
\begin{equation}
\label{eq:ne}
\frac{1}{n^2} = \frac{1}{n_e^2} + \left( \frac{1}{n_0^2} - \frac{1}{n_e^2} \right) \cdot \cos^2\theta
\end{equation}

The optical axis is given by the optical axis of medium 1 for the reflected and of medium 2 for the refracted ray. Rewriting $\cos(\theta$) as scalar product between the wave vector and the optical axis gives:
\begin{equation}
\frac{1}{n_e^2} + \left( \frac{1}{n_o^2} - \frac{1}{n_e^2}  \right) \cdot \frac{(a_x r_x + a_y r_y)^2}{n^2}- \frac{1}{n^2} = 0 .  
\end{equation}

\noindent Here $n^2=\vec{r}^2=r_x^2+r_y^2$. The solution is given as:
\begin{equation}
r_y = \frac{- \beta a_x a_y r_x \pm \sqrt{D}}{1+ \beta a_y^2}
\end{equation}

\noindent with:
\begin{equation}
\label{disc}
\begin{split}
D =  (\beta a_x a_y r_x)^2 - (1+ \beta a_y^2) (r_x^2 (1+\beta a_x^2) -n_e^2) 
 = n_e^2 \cdot (1+\beta a_y^2) - r_x^2 \cdot (1+\beta \cdot (a_x^2 + a_y^2))
\end{split}
\end{equation}

Of the two solutions the direction appropriate for the reflected/refracted ray is chosen and the other discarded. In the case of no birefringence ($\beta=0$) we again obtain the solution for the ordinary ray.

\subsection{Poynting vectors}

Once the wave vector directions are determined, the boundary continuity conditions can be written for normal components of $\vec{D}$ and $\vec{B}$, and for tangential components of $\vec{E}$ and $\vec{H}$. If $\vec{n}$ is a normal vector perpendicular to the interface surface, we have:
\begin{equation}
\vec{n} \cdot \vec{D_1} = \vec{n} \cdot \vec{D_2}, \hspace{3mm} 
\vec{n} \cdot \vec{B_1} = \vec{n} \cdot \vec{B_2}, \hspace{3mm} 
\vec{n} \times \vec{E_1} = \vec{n} \times \vec{E_2}, \hspace{3mm} 
\vec{n} \times \vec{H_1} = \vec{n} \times \vec{H_2} 
\label{eq:boundary}
\end{equation}

Here $1$ indicates the total sum of fields for incident and reflected waves, and $2$ indicates the fields of the refracted waves propagating away from the boundary surface in the second medium. Since $\vec{B}=\vec{H}$ two of the equations above simply imply that $\vec{B_1}=\vec{B_2}$ and $\vec{H_1}=\vec{H_2}$. Together with the boundary conditions for $\vec{D}$ and $\vec{E}$, this is a system of 6 linear equations. These equations are sufficient to determine the amplitudes of 4 outgoing waves: two reflected (ordinary and extraordinary), and two refracted (also, ordinary and extraordinary). Since we only have 4 unknowns, 2 of these equations are necessarily co-linear to the rest, if the wave vectors were determined  correctly.

Solving the linear equation system yields the Poynting vectors and as such the energy flux directions / photon directions of the (up to) four outgoing rays. The relative intensity of these rays, as usually denoted in Fresnel coefficients, is derived from the Poynting theorem, which for our case (no moving charges, no temporal change in total energy) is given as 
\begin{equation}
 \oiint\limits_{\partial V} \vec{S}\cdot d \vec{A} =0
\end{equation}
where $\partial V$ is the boundary of a volume V surrounding the interface. The choice of volume is arbitrary. A simple choice is a box around the interface. In the limit of an infinitely thin but wide box, it is evident that the sum of Poynting vector components normal to the interface plane is conserved. 

Evanescent waves, i.e. waves with a complex wave vector, which decay away from the boundary surface and arise when the discriminant in the wave vector equation (Eq.\ \ref{disc}) is negative, will necessarily yield vanishing contributions to such sum. As the photon interacts with a boundary there is a brief flow of energy along the surface boundary within evanescent solutions (if any), but no energy flows away from the boundary within such solutions. The evanescent waves need to be considered when solving the boundary conditions as given in equation \ref{eq:boundary}. 

The outgoing photon is then chosen randomly, with probabilities proportional to the (up to) four normal components of the non-evanescent Poynting vectors.

After implementing the solution presented here, we came across this paper  \cite{ZHANG1996549} and found the approach described there to be similar to ours. Based on the calculations above, a photon propagation simulation for birefringent polycrystals was implemented in C++.

\begin{figure}[h]
    \centering
    \subfloat{\includegraphics[width=0.25\textwidth]{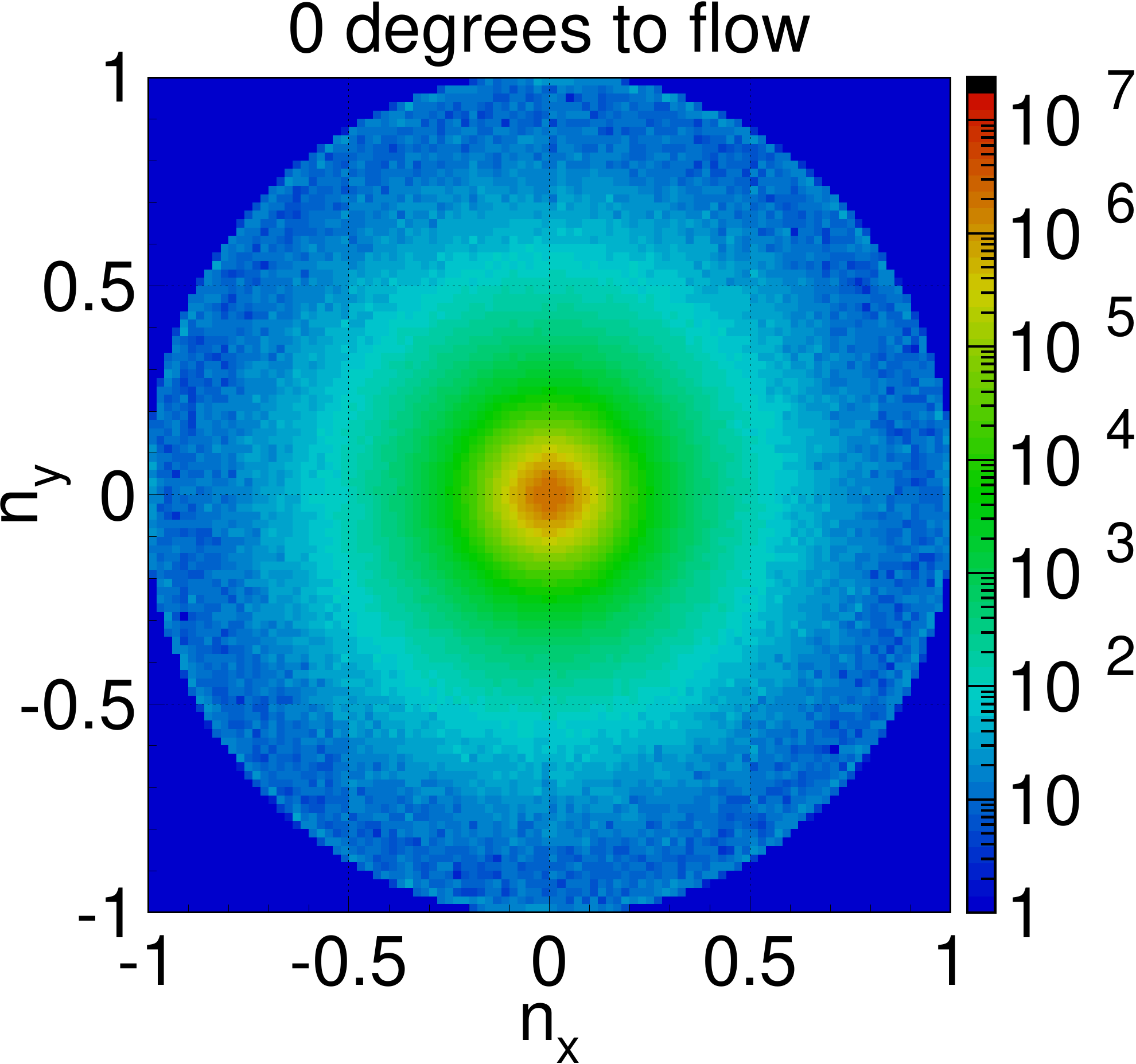}}
    \subfloat{\includegraphics[width=0.25\textwidth]{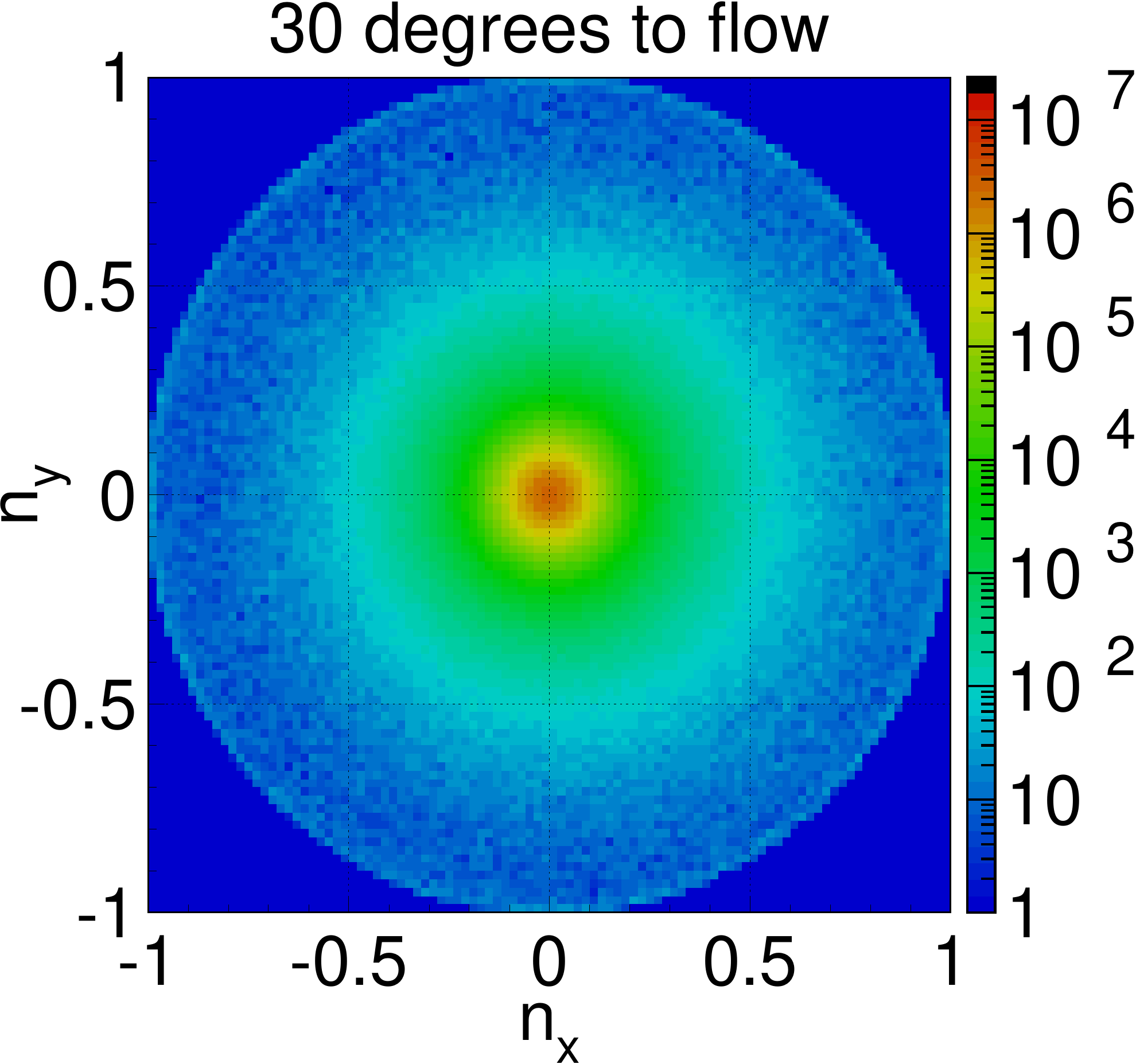}}
    \subfloat{\includegraphics[width=0.25\textwidth]{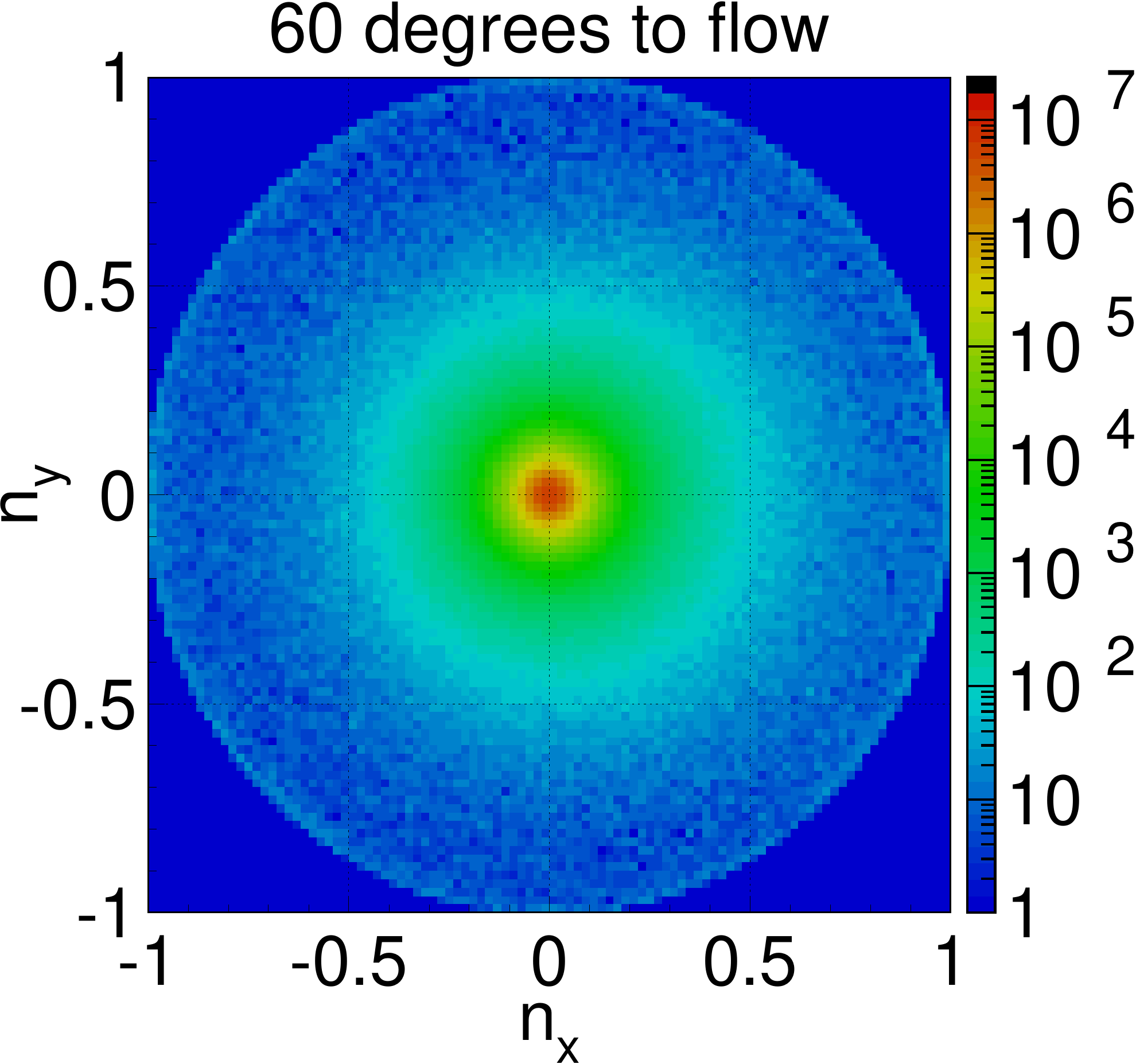}}
    \subfloat{\includegraphics[width=0.25\textwidth]{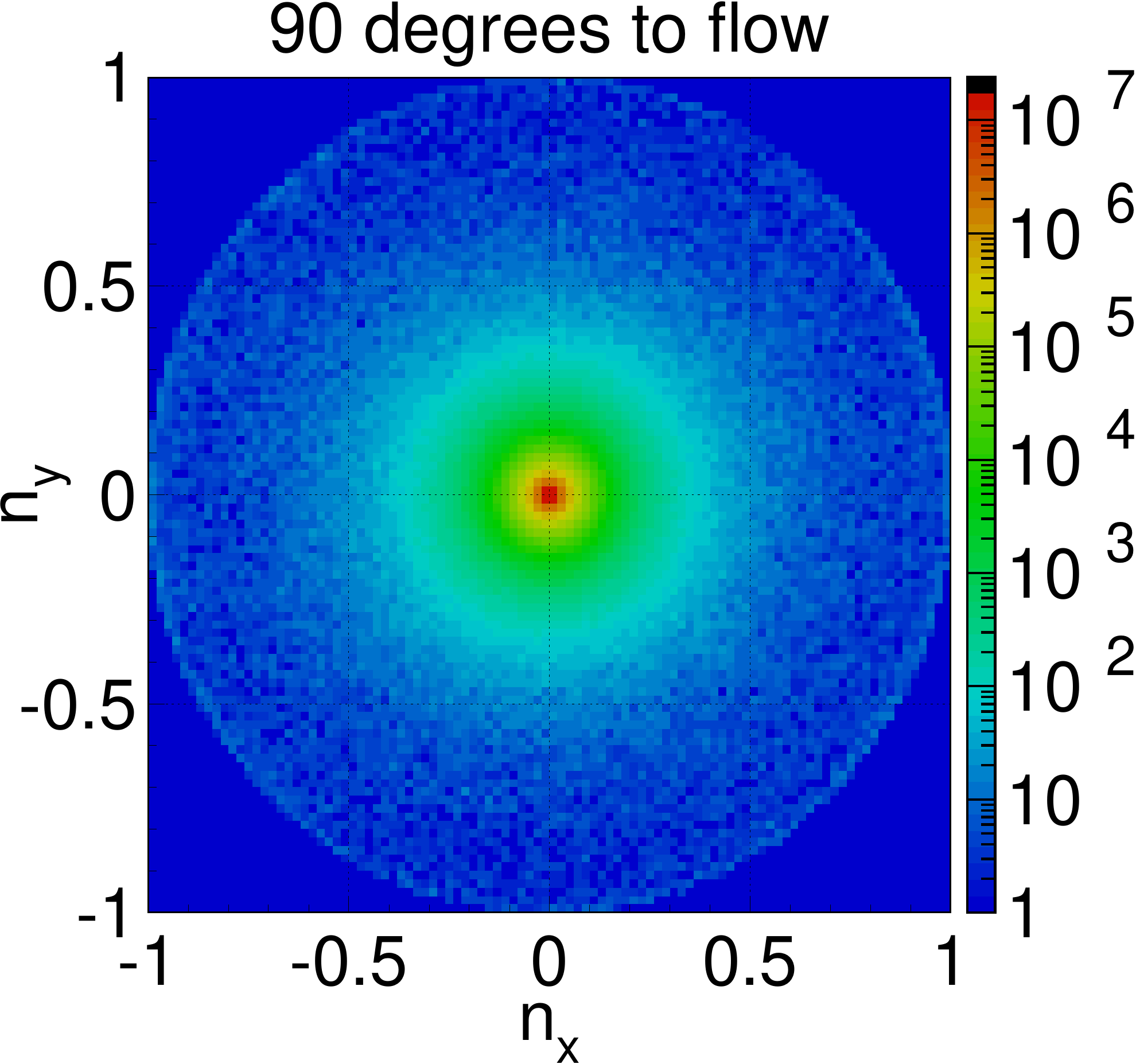}}
    \caption{Photon direction spread after propagating through 1000 grains with a perfect girdle distribution of c-axis orientations. Left: Propagating along the flow. Right: Propagating along the tilt direction. Observe the change in diffusion and the slight asymmetry for intermediate angles.}
    \label{fig:diffusionpatterns}
\end{figure}

\begin{figure}[!h]
    \centering
    \subfloat{\includegraphics[width=0.45\textwidth]{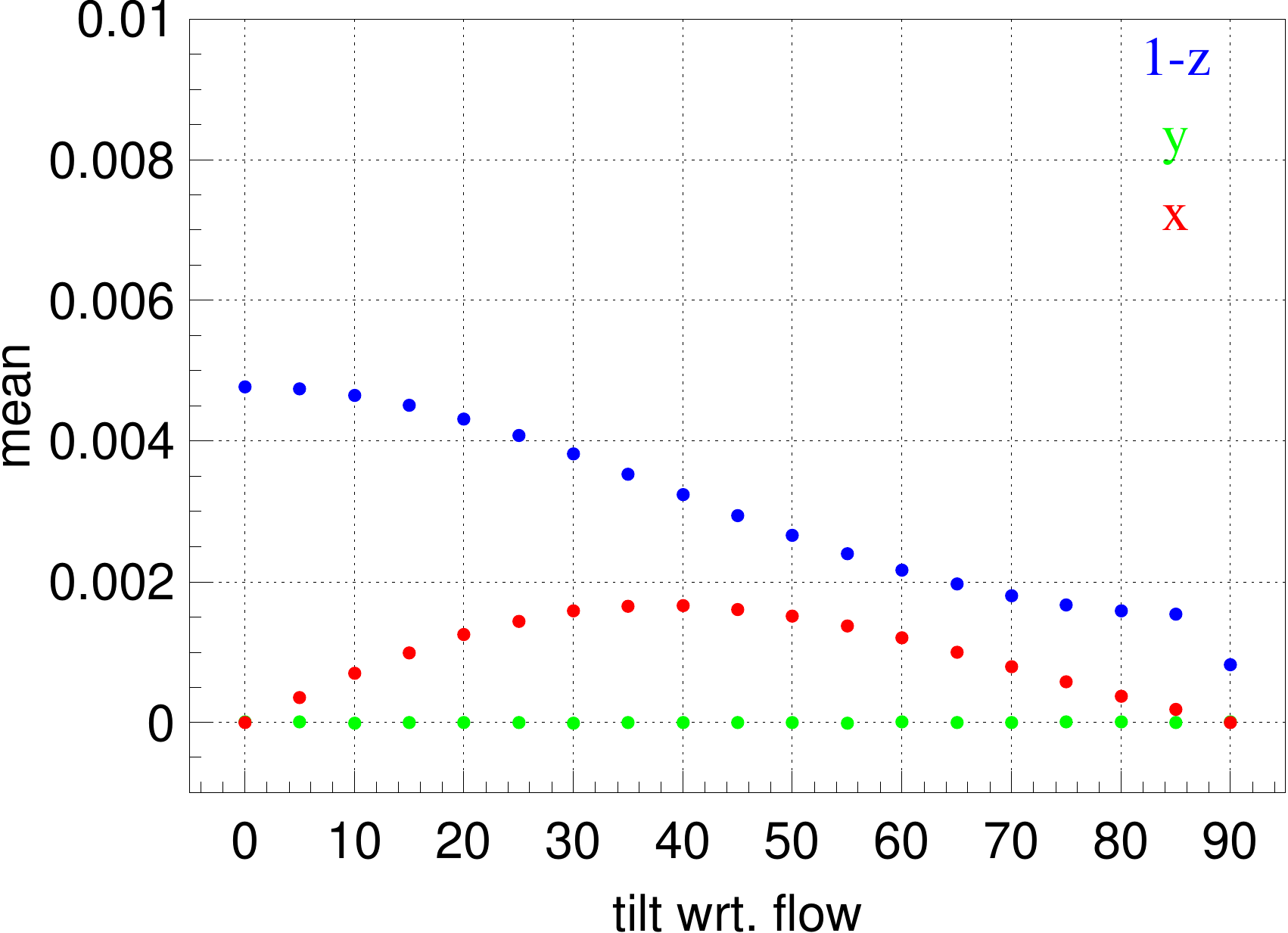}}\hfill
    \subfloat{\includegraphics[width=0.45\textwidth]{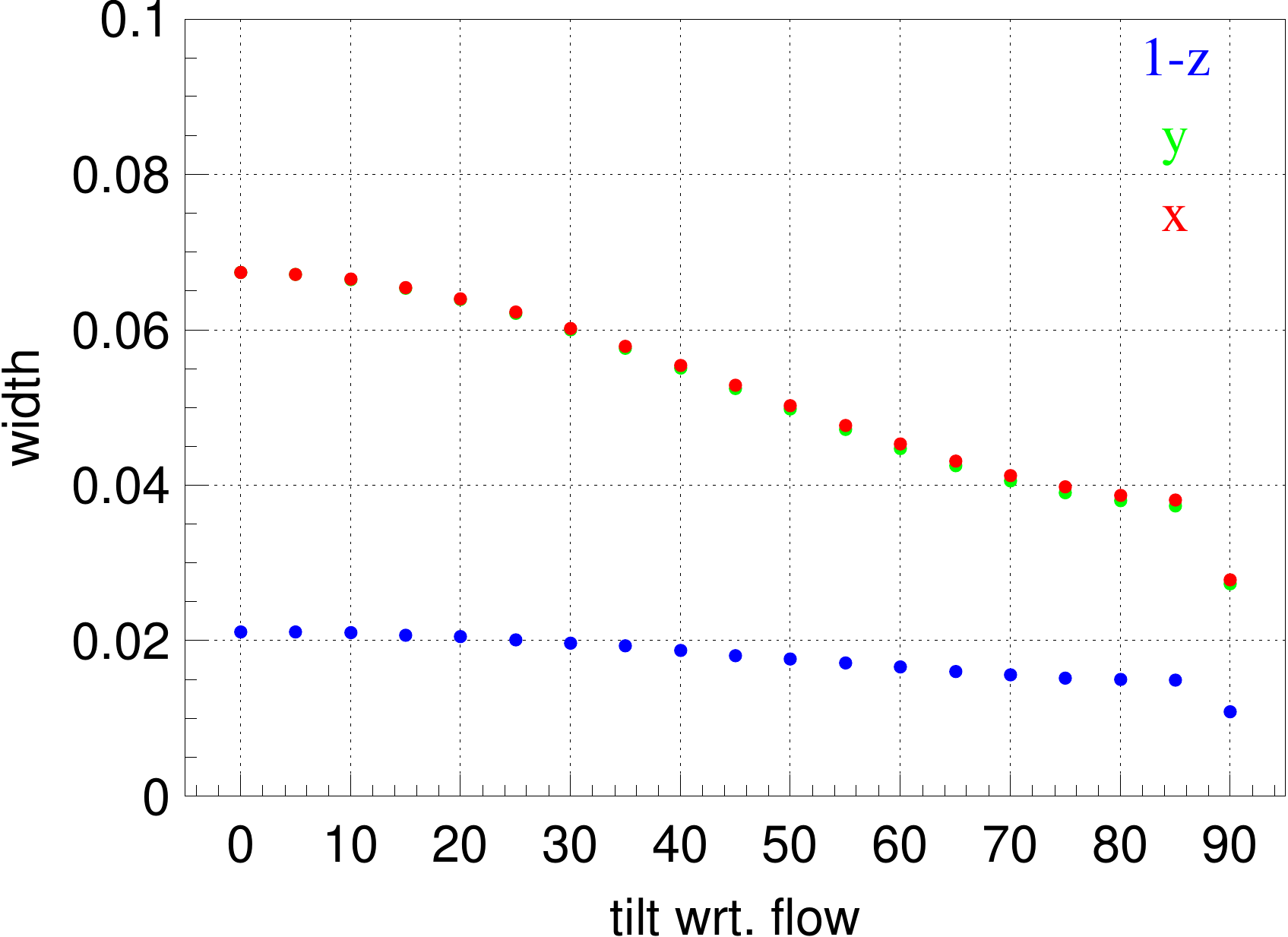}}
    \caption{Average deflection and width of the diffusion patterns for the case above. Initial photon direction is along the z-axis.}
    \label{fig:bfraverages}
\end{figure}

\section{Simulated diffusion patterns}

The resulting diffusion patterns depend on the assumed fabric and the orientation distribution of grain boundary planes encountered by the photons. The average shape and thus also the surface orientation density of ice crystals can be approximated by a triaxial ellipsoid. 

For a generalized ellipsoid the diffusion patterns are thus not only a function of the opening angle between the initial photon direction and the flow, but depend on the absolute zenith and azimuth orientation of the propagation direction with respect to the flow.

Simulated diffusion patterns after 1000 boundary crossings for a variety of initial propagation directions relative to the flow axis and assuming a spherical average grain as well as a perfect girdle fabric are shown in Figure \ref{fig:diffusionpatterns}. The overall diffusion is largest when propagating along the flow direction and gets continuously smaller towards the tilt direction. For intermediate angles the distribution is slightly asymmetric, resulting in a mean deflection towards the ice flow axis.
The averages and standard deviations of the propagation direction components for the scenario in Figure \ref{fig:diffusionpatterns} are shown in Figure \ref{fig:bfraverages}. The maximum deflection is $\sim 0.1^{\circ}$. This is on the order of magnitude which is required to describe the in-situ effect.

\section{Summary and Outlook}

While light diffusion in birefringent polycrystals is a well established effect, this report presents the first exact calculation and simulation of the resulting diffusion patterns.

In a girdle fabric, as present for most of the IceCube depths, the diffusion is largest for light propagating along the flow axis and smallest for light propagating along the tilt axis. In addition a small effective deflection towards the flow axis is experienced by photons not propagating on the flow or the tilt axis. 

The contribution of the diffusion to the bulk scattering in the ice, as well as a potential description of the optical anisotropy through the deflection remain to be studied in more detail.

\bibliographystyle{ICRC}
\bibliography{sample}
\end{document}